\documentclass[prx,twocolumn,superscriptaddress,floatfix,longbibliography]{revtex4-2}
\usepackage{enumerate}
\usepackage{graphicx}
\usepackage{amsmath}
\usepackage{natbib}
\usepackage{mathtools}
\usepackage{xcolor}
\usepackage{amsfonts}
\usepackage{amssymb}
\usepackage{units}
\usepackage[normalem]{ulem}
\newcommand{\ket}[1]{| #1 \rangle}
\newcommand{\bra}[1]{\langle #1 |}

\begin{document}
\pdfoutput=1
\today
\title{Magnon-magnon entanglement and its detection in a microwave cavity}

\author{Vahid Azimi Mousolou\footnote{Electronic address: v.azimi@sci.ui.ac.ir}}
\affiliation{Department of Applied Mathematics and Computer Science, 
Faculty of Mathematics and Statistics, 
University of Isfahan, Isfahan 81746-73441, Iran}
\affiliation{Department of Physics and Astronomy, Uppsala University, Box 516, 
SE-751 20 Uppsala, Sweden}

\author{Yuefei Liu}
\affiliation{Department of Applied Physics, School of Engineering Sciences, 
KTH Royal Institute of Technology, AlbaNova University Center, SE-10691 Stockholm, 
Sweden}

\author{Anders Bergman}
\affiliation{Department of Physics and Astronomy, Uppsala University, Box 516, 
SE-751 20 Uppsala, Sweden}

\author{Anna Delin}
\affiliation{Department of Physics and Astronomy, Uppsala University, Box 516, 
SE-751 20 Uppsala, Sweden}
\affiliation{Department of Applied Physics, School of Engineering Sciences, 
KTH Royal Institute of Technology, AlbaNova University Center, SE-10691 Stockholm, 
Sweden}
\affiliation{Swedish e-Science Research Center (SeRC), KTH Royal Institute of Technology, 
SE-10044 Stockholm, Sweden}

\author{Olle Eriksson}
\affiliation{Department of Physics and Astronomy, Uppsala University, Box 516, 
SE-751 20 Uppsala, Sweden}
\affiliation{School of Science and Technology, \"Orebro University, SE-701 82, 
\"Orebro, Sweden}

\author{Manuel Pereiro}
\affiliation{Department of Physics and Astronomy, Uppsala University, Box 516, 
SE-751 20 Uppsala, Sweden}

\author{Danny Thonig}
\affiliation{School of Science and Technology, \"Orebro University, SE-701 82, 
\"Orebro, Sweden}

\author{Erik Sj\"oqvist\footnote{Electronic address: 
erik.sjoqvist@physics.uu.se}}
\affiliation{Department of Physics and Astronomy, Uppsala University, 
Box 516, SE-751 20 Uppsala, Sweden}

\begin{abstract}
Quantum magnonics is an emerging research field, with great potential for applications in magnon based hybrid systems and quantum information processing. Quantum correlation, such as  entanglement, is a central resource in many quantum information protocols that naturally comes about in any study toward quantum technologies. This applies also to quantum magnonics. Here, we investigate antiferromagnets in which sublattices with ferromagnetic interactions can have two different magnon modes, and we show how this may lead to experimentally detectable bipartite continuous variable magnon-magnon entanglement. The entanglement can be fully characterized via a single squeezing parameter, or, equivalently, entanglement parameter. The clear relation between the entanglement parameter and the Einstein, Podolsky, and Rosen (EPR) function of the ground state opens up for experimental observation of magnon-magnon continuous variable entanglement and EPR non-locality.  We propose a practical experimental realization to detect the EPR function of the ground state, in a setting that relies on magnon-photon interaction in a microwave cavity.
\end{abstract}
\maketitle
\section{Introduction}
Hybrid quantum systems provide a natural flexible platform for quantum technologies.  
Recent developments in quantum magnon spintronics suggest that hybrid quantum systems based on collective spin-wave excitations in magnetic materials, i.e., magnons, are highly promising for many short and long terms applications in quantum technologies, including quantum sensing, quantum communication, quantum simulation, and quantum computing \cite{lachance-quirion2019, clerk2020, awschalom2021, zhang2020}. The spin-wave magnon modes interact coherently with microwave and optical photons, phonons, and superconducting qubits, which is essential for engineering efficient hybrid quantum technologies \cite{lachance-quirion2019, clerk2020}.

Recently, studies of entanglement between different components of a hybrid magnonic system have been under focus, as a fundamental element in a quantum device design \cite{lachance-quirion2019, clerk2020, li2018a, li2019, zhang2019, bossini2019, yuan2020a, yuan2020b, azimi-mousolou2020}. Relevant for the present investigation, it was suggested in Ref.~\cite{azimi-mousolou2020} that one may identify a hierarchy of magnon mode entanglement in antiferromagnetic materials. The motivation for studies of entanglement in magnetic materials can be traced to the development of modern synthetic routes that allow most combinations of elements to form a magnetic material with tailored properties. This development in materials growth, in combination with advanced lithographic techniques for nano-structuring, open up for new vistas to be explored in magnetic nano-technology \cite{lenk2011}, where magnon-magnon entanglement is an essentially unexplored research field.

Here, we discuss how an antiferromagnetic coupling between two ferromagnetic spin lattices create bipartite continuous-variable entanglement between the two ferromagnetic magnon modes in a way that each energy eigenstate of the system becomes a two-mode coherent state with nonzero entropy of entanglement. We show that the entanglement entropy of the energy eigenstates are all given by a single squeezing parameter, which can be related and measured through the Einstein, Podolsky, and Rosen (EPR) function of the ground state of the system. We also propose a feasible setup for experimental measurement of the EPR function and consequently the bipartite ferromagnetic magnon-mode entanglement and EPR non-locality. The measurement setup is based on magnon-photon coupling in a microwave cavity, which is appropriate for a wide range of antiferromagnetic materials \cite{cullity2009}
or synthetic antiferromagnetic multilayers\cite{stohr2007}.
There are many compounds with antiferromagnetism, where oxides is a broad class. In the perovskite structure alone, there are three types of antiferromagnetic structures that are relevant for the discussion presented here; A-, C and G-type antiferromagnetism \cite{solovyev2002,tokura2000}. Synthetically grown multilayers that have ferromagnetic coupling within one layer and antiferromagnetc coupling between the layers are also plentiful in the literature, and form in fact the basis for the giant magnetoresistance effect \cite{fert2008}.
 
\section{Essential aspects of magnon-magnon entanglement}
\label{MME-short}
\begin{figure}[t]
\begin{center}
\includegraphics[width=55mm]{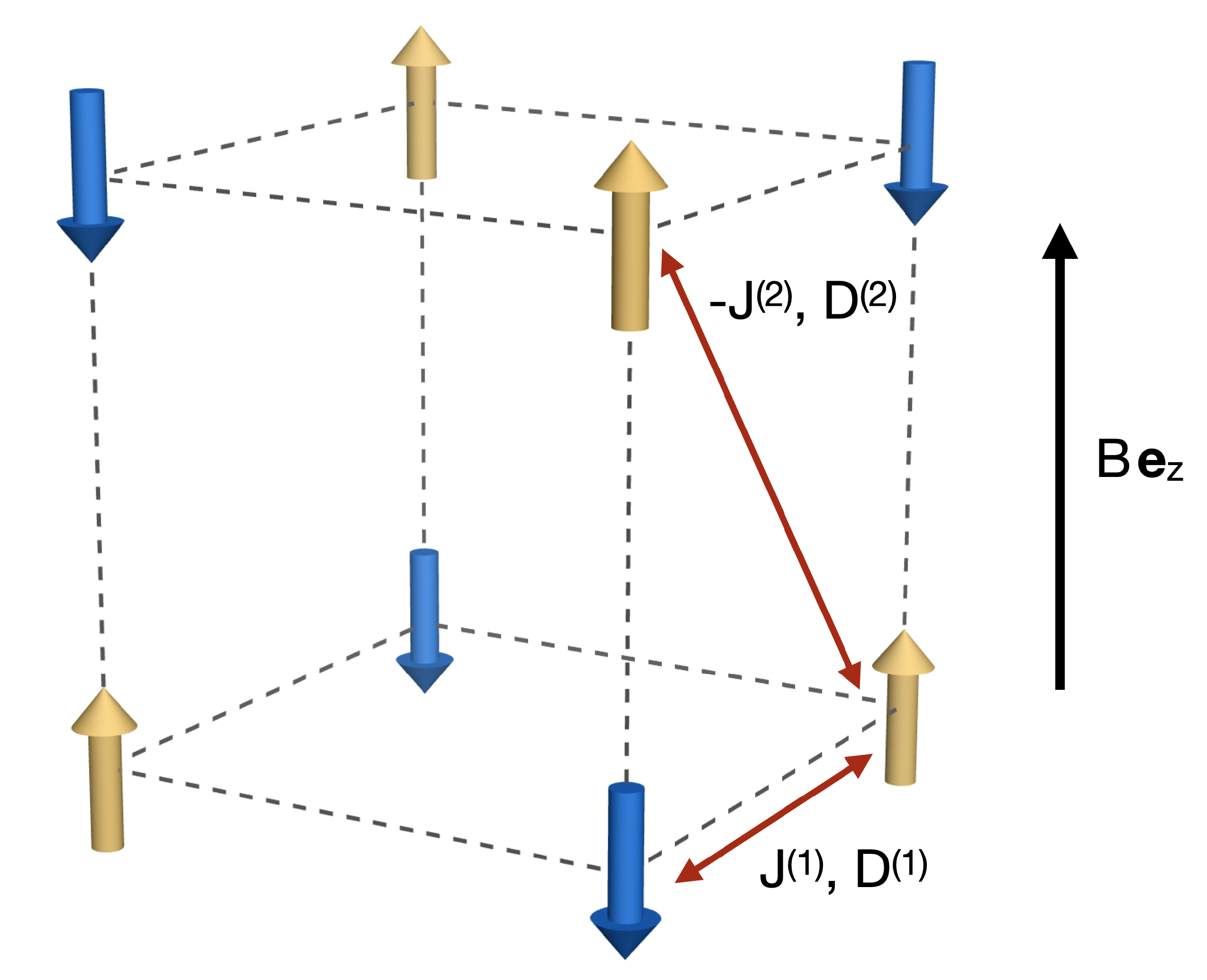}
\end{center}
\caption{(Color online) Schematic illustration of a bipartite system consist of two ferromagnetic spin sublattices with opposite magnetizations, which are coupled antiferromagnetically to each other. The whole system can be viewed as a single antiferromagnetic compound with nearest and next nearest neighbours interactions given by Heisenberg and Dzyaloshinsky-Moriya exchange interactions. The superfix $(1)$ denote couplings between the two different sublattices while $(2)$ describe couplings within the same sublattice.}
\label{fig:model}
\end{figure}

 In this section we describe briefly the Hamiltonian considered in this work, and the main conclusion as regard to which magnon modes that are most suitable for detection in an experimental setup. A full account of the technical aspects of this discussion can be found in the appendices. 
We consider a bipartite system of two ferromagnetic sublattices with opposite magnetizations denoted here by $A$ and $B$, which are coupled antiferromagnetically to each other. Figure \ref{fig:model} illustrates our spin model system for a G-type antiferromagnet structure, which is consistent with our formulation for the Hamiltonian in Eq.~\eqref{DMI}. However, the following discussion is general and holds for every bipartite antiferromagnetic structure including C and A-types, where the latter also allows for describing synthetically grown antiferromagnets. 

The magnetic interactions of the G-type system are described by the Hamiltonian 
\begin{eqnarray}
H=H_{1}+H_{2}+H_{a}+H_{z},
\label{MH}
\end{eqnarray}
where
\begin{eqnarray}
 H_{1}&=&\sum_{\langle ij\rangle}\left[J^{(1)}{\bf S}_{i} \cdot {\bf S}_{j}+ \mathbf{D}^{(1)}_{ij}\cdot\left( 
\mathbf{S}_{i}\times\mathbf{S}_{j}\right)\right],\nonumber\\
H_{2}&=&\sum_{\langle\langle ij\rangle\rangle}\left[-J^{(2)}{\bf S}_{i} \cdot {\bf S}_{j}+ \mathbf{D}^{(2)}_{ij}\cdot\left( 
\mathbf{S}_{i}\times\mathbf{S}_{j}\right)\right],\nonumber\\
 H_{a}&=&-\mathcal{K}\sum_{i}(S_{i}^{z})^{2},\ \ \ \ \ \ \ \ \ \ \ \ \ \ \mathcal{K}>0,\nonumber\\
H_{z}&=&\sum_{i} \mathbf{B}\cdot{\bf S}_{i},\ \ \ \ \ \ \ \ \ \ \ \ \ \ \ \ \ \ \mathbf{B}=B\mathbf{e}_z,\nonumber\\
J^{(1)}&>&0,  \ \ \ \ \ \   \mathbf{D}^{(1)}_{ij}=-\mathbf{D}^{(1)}_{ji}=\mathbf{D}^{(1)}=D^{(1)}\mathbf{e}_z,\nonumber\\
J^{(2)}&>&0,  \ \ \ \ \ \   \mathbf{D}^{(2)}_{ij}=-\mathbf{D}^{(2)}_{ji}=\mathbf{D}^{(2)}=D^{(2)}\mathbf{e}_z.
\label{DMI}
\end{eqnarray}
$H_{1}$ defines antiferromagnetic interaction between neighbouring sites on opposite sublattices, e.g., nearest-neighbour spins in G-type antiferromagnets; $H_{2}$ defines ferromagnetic interaction between neighbouring sites within each sublattices, e.g.,  next nearest-neighbour spins in G-type antiferromagnets.
In our model, we assume that the interaction can be described by using a Heisenberg (symmetric) exchange term combined with a Dzyaloshinsky-Moriya (DM) (antisymmetric) exchange term. $H_{a}$ is the easy axis anisotropy, and $H_{z}$ represents the Zeeman term in the Hamiltonian. Both $H_{a}$ and $ H_{z}$ are taken to be in the $z$ direction for each spin regardless of  sublattice. 

At low temperatures ($k_{B}T\ll \text{min}\{J^{(1)}, J^{(2)}\}$), one may bosonize the Hamiltonian in terms of collective modes in $\mathbf{k}$-space to arrive at 
\begin{eqnarray}
H_{\mathbf{k}} & = & 
\omega_{a_{\mathbf{k}}}a_{\mathbf{k}}^{\dagger} a_{\mathbf{k}} + 
\omega_{b_{-\mathbf{k}}}b_{-\mathbf{k}}^{\dagger} b_{-\mathbf{k}}
\nonumber \\ 
 & & + g_{\mathbf{k}} a_{\mathbf{k}} b_{-\mathbf{k}} + g^{*}_{\mathbf{k}}a_{\mathbf{k}}^{\dagger}b_{-\mathbf{k}}^{\dagger}. 
\label{MHM}
\end{eqnarray}
Here, $a_{\mathbf{k}}^{\dagger}$ ($a_{\mathbf{k}}$) and $b^{\dagger}_{-\mathbf{k}}$ ($b_{-\mathbf{k}}$) are bosonic creation (annihilation) operators, which commute and define independent families of bosonic operators on the opposite sublattices 
$A$ and $B$, respectively. Furthermore, $g_{\mathbf{k}}$ and $\omega_{a_{\mathbf{k}}}$ ($\omega_{b_{-\mathbf{k}}}$) are defined in Eq.~\eqref{CCab}.

Under SU(1,1) Bogoliubov transformation,
it is possible to obtain the following diagonal form of the Hamiltonian
 \begin{eqnarray}
H_{\mathbf{k}} = 
\omega_{\alpha_{\mathbf{k}}}\alpha_{\mathbf{k}}^{\dagger}\alpha_{\mathbf{k}} +
\omega_{\beta_{-\mathbf{k}}}\beta_{-\mathbf{k}}^{\dagger} \beta_{-\mathbf{k}}.
\label{DHH}
\end{eqnarray}
For details, see derivation of Eq.~\eqref{DHH-A} in the appendix. 
Based on these considerations, a hierarchy of magnon mode entanglement in antiferromagnets of the kind discussed here, has been analyzed in detail \cite{azimi-mousolou2020}. Here, we make use of the fact that entanglement entropies of all energy eigenbasis states in the $(a, b)$ modes can be written as
\begin{eqnarray}
E\left[(\alpha^{\dagger}_{\mathbf{k}})^{x}(\beta^{\dagger}_{-\mathbf{k}})^{y}\ket{\psi_{0}}\right]&=&-\sum_{n=0}^{\infty}|p^{(x, y)}_{n; \mathbf{k}}|^{2}\log|p^{(x, y)}_{n; \mathbf{k}}|^{2}\nonumber\\
\label{EEE}
\end{eqnarray}
where $|p^{(x, y)}_{n; \mathbf{k}}|$ are only functions of the single squeezing parameter $r_{\mathbf{k}}$ and $\ket{\psi_{0}}$ is the ground state of the system. Detailed analyses in section \ref{MME-long} of the appendix and also in Ref.~\cite{azimi-mousolou2020} show that the entanglement parameter $r_{\mathbf{k}}$ and therefore the entanglement entropies are only given by absolute value of the ratio $\Gamma_{\mathbf{k}}\propto\frac{g_{\mathbf{k}}}{\omega_{\alpha_{\mathbf{k}}}+\omega_{\beta_{-\mathbf{k}}}}$.

In the absence of an antiferromagnetic coupling term $H_{1}$ in the Hamiltonian in Eqs.~\eqref{MH}, the $(a, b)$ and $(\alpha, \beta)$ modes coincide and as a result the bosonized Hamiltonians in Eqs.~\eqref{MHM} and \eqref{DHH} would be exactly the same describing two separate non-interacting ferromagnets with opposite magnetizations, e.g., aligned in the $z$-direction so that $\langle {\bf S}_{i\in A}\rangle=-\langle {\bf S}_{j\in B}\rangle=(0, 0, S)$. In this case, the full energy spectra are separately given by the ferromagnetic bosonic modes of $a$ and $b$.
Hence, $H_{1}=0$ corresponds to a pair of disentangled ferromagnets with separable two-magnon-mode eigenstates. 
As soon as the antiferromagnetic interaction between the two ferromagnetic sublattices is turned on, i.e., $H_{1}\ne 0$, the $(\alpha, \beta)$ modes become hybridized antiferromagnetic magnon modes while $(a, b)$ still represents a pair of ferromagnetic magnon modes quantizing the two ferromagnet sublattices. Indeed, finite entanglement entropies between $a$ and $b$ magnon modes in the case of $H_{1}\ne 0$, indicate that the antiferromagnetic interaction creates bipartite continuous-variable entanglement between the two ferromagnets in a way that each energy eigenstate becomes an entangled two-magnon-mode coherent state (see section \ref{MME-long} of the appendix as well as Ref.~\cite{azimi-mousolou2020}). Each of the energy eigenstates is a coherent superposition of joint excitations of ferromagnetic magnons in $a$ and $b$ modes. 

The main focus of the work presented here is to provide a theoretical foundation, that establishes an experimental platform to measure entanglement between magnon modes of an antiferromagnetic material, as described by a pair of bosonic modes $(a, b)$ through the EPR type of nonlocality. The purely analytical work is combined with a suggested experimental setup, that relies on the magnon-photon interaction, with which realistic experimental investigations can be made.
This hence provides a formal, as well as practical, platform for which magnon entanglement may be detected experimentally. 

\section{measurement scheme via microwave cavity}

A highly relevant concept to continuous variable entanglement is the Bell type nonlocal correlations known as Einstein, Podolsky and Rosen (EPR) nonlocality. In the present case, the EPR nonlocality can be quantified by the following EPR function \cite{fadel2020}
\begin{eqnarray}
\Delta(\psi) =   
\frac{1}{2}[\text{Var}_{\psi}(X_{\mathbf{k}}^{A}+X_{\mathbf{k}}^{B})
 + \text{Var}_{\psi}(P_{\mathbf{k}}^{A}-P_{\mathbf{k}}^{B})],
\nonumber \\ 
 \label{EPRR}
\end{eqnarray}
where $X_{\mathbf{k}}^{A} = \frac{a_{\mathbf{k}} + a_{\mathbf{k}}^{\dagger}}{\sqrt{2}}$ $\left(X_{\mathbf{k}}^{B} = 
\frac{b_{\mathbf{k}}+b_{\mathbf{k}}^{\dagger}}{\sqrt{2}}\right)$ and 
$P_{\mathbf{k}}^{A} = \frac{a_{\mathbf{k}} - a_{\mathbf{k}}^{\dagger}}{i\sqrt{2}}$ $\left(P_{\mathbf{k}}^{B} = \frac{b_{\mathbf{k}} - 
b_{\mathbf{k}}^{\dagger}}{i\sqrt{2}}\right)$ are assumed to be the dimensionless 
position and momentum quadratures for the $a_{\mathbf{k}} (b_{\mathbf{k}})$ mode, respectively. The 
$\text{Var}_{\psi}(V)$ is the variance of an Hermitian operator $V$ with respect to the state $\ket{\psi}$.
The uncertainty relation $\Delta(\psi)\ge 1$ is known to hold for any given bipartite separable state $\ket{\psi}$  \cite{fadel2020}.  Therefore, any violation of this inequality is an indication of the state $\ket{\psi}$ being nonlocal and indeed a bipartite entangled state.
Note that the EPR nonlocality specifies a stronger type of entanglement than a nonzero entropy of entanglement in a sense that there are states with nonzero entropy of entanglement which do not violate the uncertainty relation. Note also that the EPR nonlocality depends, just like the entanglement entropy, on the modes that are chosen. In the present discussion we consider $(a, b)$ modes.

\begin{figure}[h]
\begin{center}
\includegraphics[width=80mm, height=40mm]{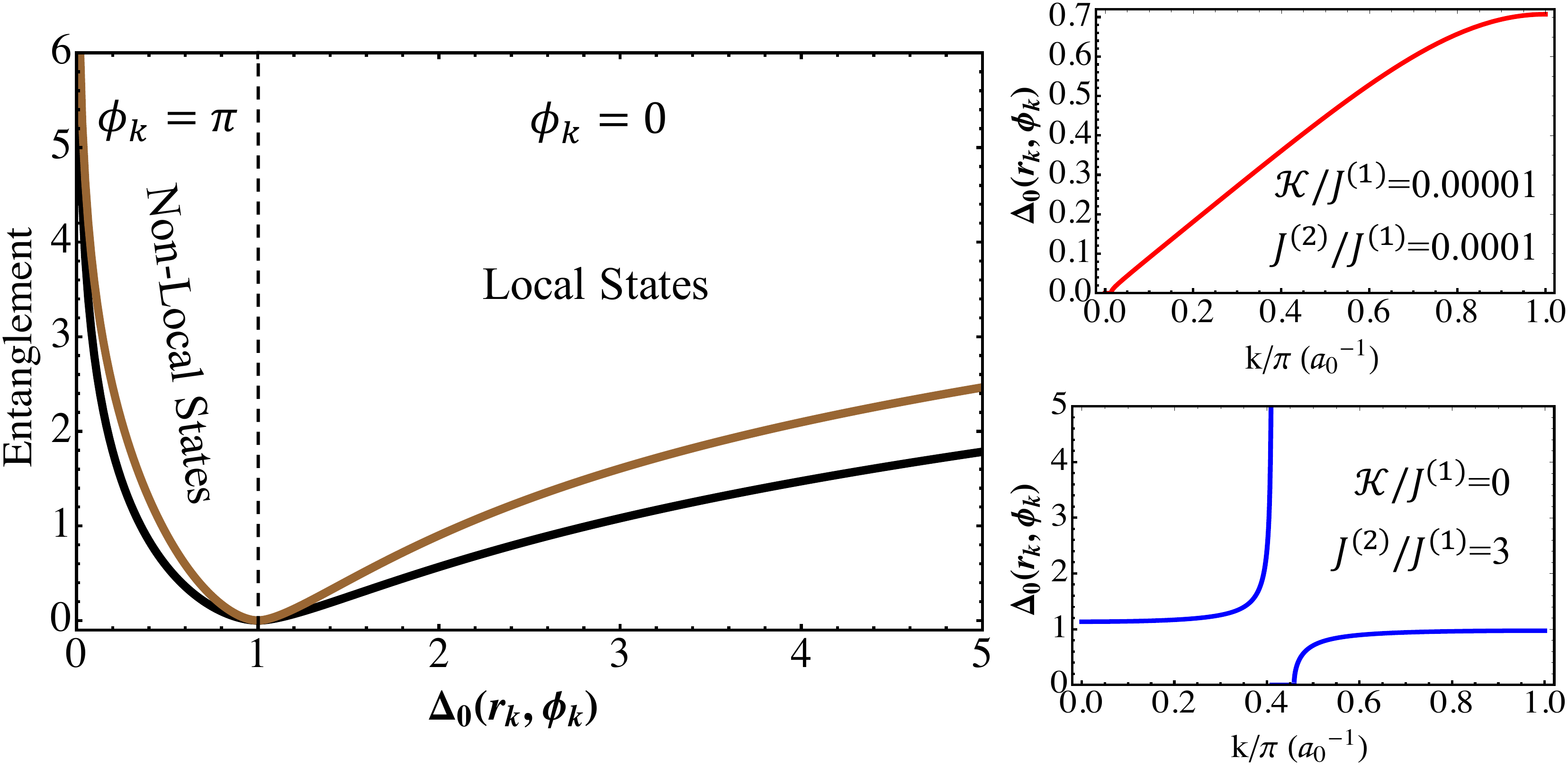}
\end{center} 
\caption{(Color online) Left panel: entanglement entropies $E[\ket{\psi_{0}(r_{\mathbf{k}}, \phi_{\mathbf{k}})}]$ and
 $E[\alpha^{\dagger}_{\mathbf{k}}\ket{\psi_{0}(r_{\mathbf{k}}, \phi_{\mathbf{k}})}]=E[\beta^{\dagger}_{-\mathbf{k}}\ket{\psi_{0}(r_{\mathbf{k}}, \phi_{\mathbf{k}})}]$ against 
the EPR function $\Delta_{0}(r_{\mathbf{k}},  \phi_{\mathbf{k}})$ for antiferromagnetic spin lattices with only Heisenberg interactions. Black and brown curves correspond to the entanglement of the ground state and the first excited states, respectively. Right panels illustrate how the EPR function $\Delta_{0}(r_{\mathbf{k}},  \phi_{\mathbf{k}})$ depends on $\mathbf{k}$ along the $(0, 0, 1)$ direction in a simple cubic lattice for selected values of exchange couplings $J^{(1)},J^{(2)}$, and magnetic anisotropy $\mathcal K$.}
\label{fig:EvsEPR}
\end{figure}

For the spin wave ground state $\ket{\psi_{0}(r_{\mathbf{k}}, \phi_{\mathbf{k}})}$ given in Eq.~\eqref{eq:GSatab}, we obtain the EPR function
\begin{eqnarray}
\Delta_{0}(r_{\mathbf{k}},  \phi_{\mathbf{k}}) = \cosh 2r_{\mathbf{k}}+\sinh2r_{\mathbf{k}}\cos\phi_{\mathbf{k}},
\end{eqnarray}
which specifies the relation between the entanglement parameter and EPR nonlocality. Since the ground state EPR nonlocality and the entanglement entropies depend on the same squeezing parameter, one may analyze the dependence between entanglement entropy and the ground state EPR nonlocality.
Figure~\ref{fig:EvsEPR} illustrates the two-mode magnon entanglement in the ground state and first excited states against the EPR function $\Delta_{0}(r_{\mathbf{k}},  \phi_{\mathbf{k}})$, for antiferromagnetic spin lattices, where only Heisenberg interactions are relevant (DM interactions are neglected) and described by Eqs.~\eqref{MH} and \eqref{DMI}. Typical materials that are known to be described by this type of spin-Hamiltonian are antiferromagnets like BiFeO$_3$ \cite{roginska1965}
and LaMnO$_3$ \cite{matsumoto1970}, as well as synthetic, antiferromagnetic multilayers \cite{fert2008}. In this case, $\Gamma_{\mathbf{k}}$ is real-valued and 
\begin{eqnarray}
\Delta_{0}(r_{\mathbf{k}},  \phi_{\mathbf{k}}) = 
\left\{ \begin{array}{ll} 
e^{2r_{\mathbf{k}}}, &\ \  \mathrm{if} \ \ \ \ \phi_{\mathbf{k}} = 0\ (\Gamma_{\mathbf{k}} < 0), \\ 
e^{-2r_{\mathbf{k}}}, &\ \  \mathrm{if} \ \ \ \ \phi_{\mathbf{k}} =\pi\ (\Gamma_{\mathbf{k}} > 0).
\end{array} \right. 
\end{eqnarray}

Figure~\ref{fig:EvsEPR} distinguishes two regions, the non-local bipartite entangled state and the local bipartite entangled state with transition point at $\Delta_{0}(r_{\mathbf{k}},  \phi_{\mathbf{k}})=1$. The clear relation between the EPR function and the two-mode magnon entanglement entropy allows for experimental detection of magnon-magnon entanglement, as we discuss in detail below. It is worth mentioning that the EPR nonlocality has been used for verification of entanglement in optical and atomic systems based on homodyne detection and types of interferometry setups \cite{gross2011, armstrong2015, peise2015, lee2016, kunkel2018, fadel2018, li2020}. However, these types of measurement setups are not realistic 
for magnon systems, since these technologies are mainly based on beam splitters that have limitations for detecting magnon entanglement. We propose as a solution, a mechanism and measurement setup that relies on light-matter interaction as a probe to 
observe the EPR function and thus EPR nonlocality and magnon-magnon entanglement.

Cavity modes can couple to magnon modes in both antiferromagnets and ferromagnets \cite{shrivastava1979, huebl2013, tabuchi2014, yuan2017, xiao2019, johansen2018}. 
The magnon-photon interaction  
can be understood in terms of the vector potential of the quantized photon-field and the magnetic moments of the material. Since the mode polarization 
and propagation direction of the electromagnetic wave are highly relevant \cite{xiao2019}, we first introduce the essential physics of this interaction. After this, a possible structure of an experimental magnon entanglement detection is discussed. 

We assume microwave cavity electromagnetic field described by the vector potential 
\begin{eqnarray}
\mathbf{A}_{\mathbf{k}}(\mathbf{r}, t)&=&\mathbf{A}_{L; \mathbf{k}}(\mathbf{r}, t)+\mathbf{A}_{R; \mathbf{k}}(\mathbf{r}, t)\nonumber\\
\mathbf{A}_{R; \mathbf{k}}(\mathbf{r}, t)&=&A_{0}\left[\mathbf{e}_{R}c_{\mathbf{k}}e^{-i(\mathbf{k} \cdot \mathbf{r}+\omega t)}+\mathbf{e}^{*}_{R}c^{\dagger}_{\mathbf{k}}e^{i(\mathbf{k} \cdot \mathbf{r}+\omega t)}\right]
\nonumber\\
&=&e^{it\omega c^{\dagger}_{\mathbf{k}}c_{\mathbf{k}}}\mathbf{A}_{R; \mathbf{k}}(\mathbf{r}, 0)e^{-it\omega c^{\dagger}_{\mathbf{k}}c_{\mathbf{k}}}
\nonumber\\
\mathbf{A}_{L; \mathbf{k}}(\mathbf{r}, t)&=&A_{0}\left[\mathbf{e}_{L}d_{-\mathbf{k}}e^{i(\mathbf{k} \cdot \mathbf{r}-\omega t)}+\mathbf{e}^{*}_{L}d^{\dagger}_{-\mathbf{k}}e^{-i(\mathbf{k} \cdot \mathbf{r}-\omega t)}\right]
\nonumber\\
&=&e^{it\omega d^{\dagger}_{-\mathbf{k}}d_{-\mathbf{k}}}\mathbf{A}_{L; \mathbf{k}}(\mathbf{r}, 0)e^{-it\omega d^{\dagger}_{-\mathbf{k}}d_{-\mathbf{k}}}\nonumber\\
\end{eqnarray}
for a given $\mathbf{k}$ vector describing the propagation direction of the electromagnetic wave. $A_{0}$ is the amplitude of the vector potential and $\omega$ is the single cavity mode frequency, which are both tuned by the volume of the cavity and the separation 
distance between the two conductor plates in the cavity (Fig.2). 
Here, we focus on the lowest energy cavity mode and disregard contributions from the higher energy cavity modes. In fact, the vector potential represents superposition of right and left circularly polarized photons, where $c_{\mathbf{k}} (c^{\dagger}_{\mathbf{k}})$ and $d_{-\mathbf{k}} (d^{\dagger}_{-\mathbf{k}})$ are the corresponding annihilation (creation) operators with unit vectors $\mathbf{e}_{R}$ and $\mathbf{e}_{L}$, respectively. In the rotating frame, the magnon-photon coupling is given by the interaction Hamiltonian 
\begin{eqnarray}
H_{\text{mp}}=H+H_{\text{ph}}-\mathbf{B}_{p}\cdot\mathbf{S},
\label{CH}
\end{eqnarray}
where $\mathbf{B}_{p}=\nabla\times \mathbf{A}_{L; \mathbf{k}}(\mathbf{r}, 0)$ is the photon induced magnetic field interacting through a Zeeman term with the total spin $\mathbf{S}$ of the antiferromagnetic material. $H$ is the spin Hamiltonian (Eq.~(1)) and 
\begin{eqnarray}
H_{\text{ph}}=\omega\sum_{\mathbf{k}}(c^{\dagger}_{\mathbf{k}}c_{\mathbf{k}}+d^{\dagger}_{-\mathbf{k}}d_{-\mathbf{k}}),
\end{eqnarray}
is the cavity photon Hamiltonian. For $\mathbf{k}$ vectors along the $(0, 0, 1)$ direction, we assume $\mathbf{e}_{R}=-\mathbf{e}^{*}_{L}=\frac{1}{\sqrt{2}}(1, -i, 0)$ and apply the Holstein-Primakoff, Fourier, and Bogoliubov transformations, as described above, to derive  
the bosonized interaction Hamiltonian  
\begin{eqnarray}
H_{\text{mp}}&=&\sum_{\mathbf{k}}H_{\text{mp}; \mathbf{k}},\nonumber\\
H_{\text{mp}; \mathbf{k}}&=&\omega_{\alpha_{\mathbf{k}}}\alpha_{\mathbf{k}}^{\dagger}\alpha_{\mathbf{k}} +
\omega_{\beta_{-\mathbf{k}}}\beta_{-\mathbf{k}}^{\dagger} \beta_{-\mathbf{k}},
+\omega(c^{\dagger}_{\mathbf{k}}c_{\mathbf{k}}+d^{\dagger}_{-\mathbf{k}}d_{-\mathbf{k}})\nonumber\\
&&+(\Delta_{\mathbf{k}}d^{\dagger}_{-\mathbf{k}}\beta_{-\mathbf{k}}+\Delta^{*}_{\mathbf{k}}d_{-\mathbf{k}}\beta^{\dagger}_{-\mathbf{k}})\nonumber\\
&&-(\Delta_{\mathbf{k}}c^{\dagger}_{\mathbf{k}}\alpha_{\mathbf{k}}+\Delta^{*}_{\mathbf{k}}c_{\mathbf{k}}\alpha^{\dagger}_{\mathbf{k}}),
\label{BCH}
\end{eqnarray}
with the resonant magnon-photon interaction. Here $\Delta_{\mathbf{k}}=\lambda_{\mathbf{k}}(u_{\mathbf{k}}+v^{*}_{\mathbf{k}})$ for $\lambda_{\mathbf{k}}=A_{0}k\sqrt{S}$ and $\mathbf{k}=(0, 0, k)$. Note that, apart from the momentum conservation which leads to Eq.~\eqref{BCH}, for the sake of energy conservation the  
off-resonant interaction $ (\Delta_{\mathbf{k}}d_{-\mathbf{k}}\alpha_{\mathbf{k}}+\Delta^{*}_{\mathbf{k}}d^{\dagger}_{-\mathbf{k}}\alpha^{\dagger}_{\mathbf{k}})-(\Delta_{\mathbf{k}}c_{\mathbf{k}}\beta_{-\mathbf{k}}+\Delta^{*}_{\mathbf{k}}c^{\dagger}_{\mathbf{k}}\beta^{\dagger}_{-\mathbf{k}})$ is  neglected.

Considering the Hamiltonian in Eq.~\eqref{BCH} for a given $\mathbf{k}$, we notice the invariant space spanned by the ordered one-particle Fock states 
\begin{eqnarray}
\ket{1}&=&\ket{1000}_{\mathbf{k}}=\alpha^{\dagger}_{\mathbf{k}}\ket{0000}_{\mathbf{k}},\nonumber\\
\ket{2}&=&\ket{0100}_{\mathbf{k}}=c^{\dagger}_{\mathbf{k}}\ket{0000}_{\mathbf{k}},\nonumber\\
\ket{1'}&=&\ket{0010}_{\mathbf{k}}=\beta^{\dagger}_{-\mathbf{k}}\ket{0000}_{\mathbf{k}},\nonumber\\
\ket{2'}&=&\ket{0001}_{\mathbf{k}}=d^{\dagger}_{-\mathbf{k}}\ket{0000}_{\mathbf{k}},
\end{eqnarray} 
with $\ket{0000}_{\mathbf{k}}$ being the joint vacuum state of the magnon-photon system. In this four dimensional invariant subspace, the Hamiltonian takes the following 
direct sum 
form 
\begin{eqnarray}
H_{\text{mp}; \mathbf{k}}=
\left(
\begin{array}{cc}
\omega_{\alpha_{\mathbf{k}}}& -\Delta^{*}_{\mathbf{k}} \\
-\Delta_{\mathbf{k}} & \omega
\end{array}
\right)\oplus
\left(
\begin{array}{cc}
\omega_{\beta_{-\mathbf{k}}}& \Delta^{*}_{\mathbf{k}} \\
\Delta_{\mathbf{k}} & \omega
\end{array}
\right).
\label{eq:DSH}
\end{eqnarray}
This implies that 
we have the following nonzero population and transition rates 
\begin{eqnarray}
|\bra{1}U(t)\ket{1}|^{2}&=&|\bra{2}U(t)\ket{2}|^{2}\nonumber\\
&=&\cos^{2}\left[t \pi f_{\mathbf{k}}\right]+\frac{(\Delta\omega_{\alpha_{\mathbf{k}}})^{2}}{(\Delta\omega_{\alpha_{\mathbf{k}}})^{2}+|\Delta_{\mathbf{k}}|^{2}}\sin^{2}\left[t \pi f_{\mathbf{k}}\right] , \nonumber\\
|\bra{1}U(t)\ket{2}|^{2}&=&|\bra{2}U(t)\ket{1}|^{2}\nonumber\\
&=&\frac{|\Delta_{\mathbf{k}}|^{2}}{(\Delta\omega_{\alpha_{\mathbf{k}}})^{2}+|\Delta_{\mathbf{k}}|^{2}}\sin^{2}\left[t \pi f_{\mathbf{k}}\right] , \nonumber\\
|\bra{1'}U(t)\ket{1'}|^{2}&=&|\bra{2'}U(t)\ket{2'}|^{2}\nonumber\\
&=&\cos^{2}\left[t \pi f'_{\mathbf{k}}\right]+\frac{(\Delta\omega_{\beta_{-\mathbf{k}}})^{2}}{(\Delta\omega_{\beta_{-\mathbf{k}}})^{2}+|\Delta_{\mathbf{k}}|^{2}}\sin^{2}\left[t \pi f'_{\mathbf{k}}\right] , \nonumber\\
|\bra{1'}U(t)\ket{2'}|^{2}&=&|\bra{2'}U(t)\ket{1'}|^{2}\nonumber\\
&=&\frac{|\Delta_{\mathbf{k}}|^{2}}{(\Delta\omega_{\beta_{-\mathbf{k}}})^{2}+|\Delta_{\mathbf{k}}|^{2}}\sin^{2}\left[t \pi f'_{\mathbf{k}}\right] , \nonumber\\
\label{TPD}
\end{eqnarray}
where $U(t)=e^{-itH_{\text{mp}}}$ is the time evolution operator and $\Delta\omega_{\alpha_{\mathbf{k}}}=\frac{\omega_{\alpha_{\mathbf{k}}}-\omega}{2}$ and $\Delta\omega_{\beta_{-\mathbf{k}}}=\frac{\omega_{\beta_{-\mathbf{k}}}-\omega}{2}$ are the magnon-photon energy differences. Transition frequencies read $\pi f_{\mathbf{k}}=\frac{\pi}{T_{\mathbf{k}}}=\sqrt{(\Delta\omega_{\alpha_{\mathbf{k}}})^{2}+|\Delta_{\mathbf{k}}|^{2}}$ and $\pi f'_{\mathbf{k}}=\frac{\pi}{T'_{\mathbf{k}}}=\sqrt{(\Delta\omega_{\beta_{-\mathbf{k}}})^{2}+|\Delta_{\mathbf{k}}|^{2}}$, which result explicitly in relations to the EPR function 
\begin{eqnarray}
\Delta_{0}(r_{\mathbf{k}},  \phi_{\mathbf{k}}) & = & \frac{(\pi f_{\mathbf{k}})^{2}-(\Delta\omega_{\alpha_{\mathbf{k}}})^{2}}{\lambda^{2}_{\mathbf{k}}}
\nonumber \\ 
 & = & \frac{(\pi f'_{\mathbf{k}})^{2}-(\Delta\omega_{\beta_{-\mathbf{k}}})^{2}}{\lambda^{2}_{\mathbf{k}}}. 
\label{EPR-TPD}
\end{eqnarray}
This, in fact, follows from 
\begin{eqnarray}
\frac{|\Delta_{\mathbf{k}}|^{2}}{\lambda^{2}_{\mathbf{k}}}=|u_{\mathbf{k}}+v^{*}_{\mathbf{k}}|^{2}&=&\cosh 2r_{\mathbf{k}}+\sinh2r_{\mathbf{k}}\cos\phi_{\mathbf{k}}\nonumber\\
&=&\Delta_{0}(r_{\mathbf{k}},  \phi_{\mathbf{k}}).
\end{eqnarray}

Equations \eqref{TPD} and\ \eqref{EPR-TPD} indicate that by tuning the photon frequency $\omega$ to either of the magnon frequencies $\omega_{\alpha_{\mathbf{k}}}$ or $\omega_{\beta_{-\mathbf{k}}}$, the EPR function and consequently the EPR nonlocality and magnon-magnon entanglement can be measured through the relevant transition frequency $f_{\mathbf{k}}$ or $f'_{\mathbf{k}}$, where the associated magnon-photon transition intensity between $1\leftrightarrow 2$ or $1'\leftrightarrow 2'$, i.e. the visibility of magnon-photon interference fringes, is maximal. From Eq.~\eqref{TPD} we note that  the 
transition frequencies are the same as the corresponding population frequencies. Therefore, the measurement of  instantaneous population rate $|\bra{x}U(t)\ket{x}|^{2}$ for each magnon or photon, where $\ket{x}$ represents the one-particle Fock states $\ket{1}, \ket{2}, \ket{1'}$ or $\ket{2'}$, also allows for detecting the EPR function.

A key practical feature is the direct sum form of the Hamiltonian in Eq.~\eqref{eq:DSH}, which results from energy and momentum conservation mentioned above. This assures that there are no transitions between the two subspaces $\mathcal{M}=\text{span}\{\ket{1}, \ket{2}\}$ and $\mathcal{M}'=\text{span}\{\ket{1'}, \ket{2'}\}$ at any time. This implies that each of the polarized photons independently interacts with its hybridized $\alpha$ or $\beta$ magnon mode counterpart. This  feature, which follows from Eq.~\eqref{eq:DSH} together with Eqs.~\eqref{TPD} and\ \eqref{EPR-TPD}, indicates that, in fact, only a single circularly polarized cavity field is needed to detect the EPR function through magnon-photon transition frequency. 

\begin{figure}[h]
\begin{center}
\includegraphics[width=85mm, height=40mm]{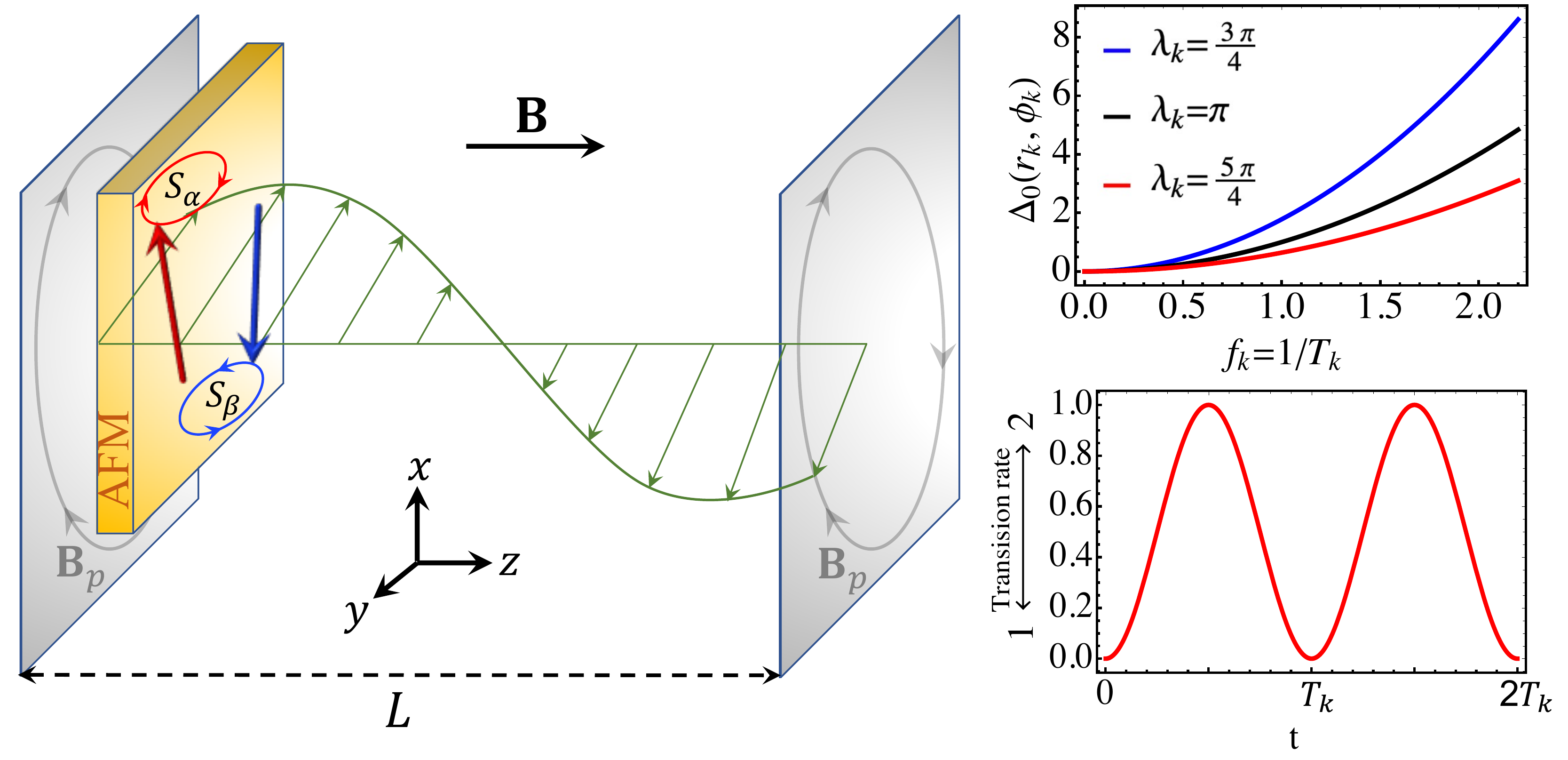}
\end{center}
\caption{(Color online) Left panel: a microwave cavity consisting of two perfect conductor plates located on the $z$ axis at a distance $L$ from each other. The hybridized magnon in an antiferromagnet (schematically shown by the yellow slab), which combines two ferromagnetic spin sublattices with opposite magnetization along the $z$ direction, is coupled to a quantized circularly polarized cavity photon field (green curve, including arrows). Right panels: the resonant magnon-photon coupling allows for measuring the EPR function $\Delta_{0}(r_{\mathbf{k}},  \phi_{\mathbf{k}})$ via the magnon-photon transition frequency $f_{\mathbf{k}}$. 
The oscillation period $T_{\mathbf{k}}$ of the magnon-photon transition rate determines the transition frequency.
An external magnetic field tunes the resonance frequency of the magnon while the photon frequency depends on the ratio between the speed of light and the separation distance $L$.}
\label{fig:experimental-setup}
\end{figure}

Figure \ref{fig:experimental-setup} (left panel) shows the measurement scheme associated with the subspace $\mathcal{M}$. In this figure, a quantized circularly polarized photon is depicted to interact with the hybridized magnon-mode ($\alpha$ mode) in a microwave cavity.  By a careful control of the magnon and photon frequencies, one can detect the EPR function through the transition frequency $f_{\mathbf{k}}=1/T_{\mathbf{k}}$ between the magnonic $\alpha$-mode and the interacting photonic mode, and therefore evaluate the bipartite continuous variable magnon-magnon entanglement. In the right panels of Figure \ref{fig:experimental-setup}, this is illustrated via the transition rate (lower right panel) between magnon and photon modes, as function of time. The oscillation period, $T_{\mathbf{k}}$, of the magnon-photon transition rate, which can be measured via a photon counting technique, determines the transition frequency $f_{\mathbf{k}}$. The figure also shows the EPR function as a function of $f_{\mathbf{k}}=1/T_{\mathbf{k}}$ (explicitly defined in Eq.~\eqref{EPR-TPD} and below) for selected values of $\lambda_{\mathbf{k}}$ (upper right panel).
A similar $\mathcal{M}'$-based setup is equally valid. 

\section{conclusion}

In conclusion, we have shown that an antiferromagnetic coupling between two ferromagnetic sublattices creates magnon-magnon entanglement in such a way that each energy eigenstate becomes a two-mode entangled state. The bipartite magnon-mode entanglement is fully characterized by a single entangling parameter, which is clearly related to the Einstein, Podolsky, and Rosen (EPR) function of the ground state. We propose a new and feasible 
measurement setup based on light and matter interaction to observe the EPR function 
through measurement of the magnon-photon transition frequency. The proposed setup is compatible with current advances in magnonic and photonic technologies. 

\section*{acknowledgments}
The authors acknowledge financial support from Knut and 
Alice Wallenberg Foundation through Grant No. 2018.0060. A.D. acknowledges 
financial support from the Swedish Research Council (VR) through Grants No. 2015-04608,  
2016-05980, and VR 2019-05304. O.E. acknowledges support from the Swedish Research Council (VR), the Swedish Foundation for Strategic Research (SSF), the Swedish Energy Agency (Energimyndigheten), ERC (synergy grant FASTCORR, project 854843), eSSENCE, and STandUPP. D.T. acknowledges support from the Swedish Research 
Council (VR) through Grant No. 2019-03666. E.S. acknowledges financial support from 
the Swedish Research Council (VR) through Grant No. 2017-03832. Some of the 
computations were performed on resources provided by the Swedish 
National Infrastructure for Computing (SNIC) at the National Supercomputer Center (NSC), 
Link\"oping University, the PDC Centre for High Performance Computing (PDC-HPC), KTH, 
and the High Performance Computing Center North (HPC2N), Ume{\aa} University.

\appendix
\numberwithin{equation}{section}
\section{Description of Hamiltonian in term of bosonic operators}

The magnetic interactions of the system are described by the Hamiltonian in Eqs.~\eqref{MH} and \eqref{DMI}, as detailed in the main text. To bosonize the Hamiltonian in terms of collective modes in ${\bf k}$-space (see Ref.~\cite{rezende2019} for a review), we use the Holstein-Primakoff transformation, 
\begin{eqnarray}
\text{Sublattice} \ A: \left\{
\begin{array}{lr}
S^{z}_{i} = S-a^{\dagger}_{i}a_{i} , 
\nonumber\\
S^{+}_{i} = \left( 2S-a^{\dagger}_{i}a_{i} \right)^{\frac{1}{2}} a_{i} , 
\nonumber\\
S^{-}_{i}=a^{\dagger}_{i} \left( 2S-a^{\dagger}_{i}a_{i} 
\right)^{\frac{1}{2}} , 
\end{array} \right.\nonumber\\
\text{Sublattice} \ B:\left\{
\begin{array}{lr}
S^{z}_{j} = b^{\dagger}_{j} b_{j} - S , 
\nonumber\\
S^{+}_{j} = b^{\dagger}_{j} \left( 2S-b^{\dagger}_{j}b_{j} 
\right)^{\frac{1}{2}} , 
\nonumber \\
S^{-}_{j} = \left( 2S-b^{\dagger}_{j}b_{j} \right)^{\frac{1}{2}} b_{j} ,
\end{array}
\right. 
\end{eqnarray}
and linear approximation at low temperatures ($k_{B}T\ll \text{min}\{J^{(1)}, J^{(2)}\}$), where $\langle a^{\dagger}_{i} a_{i} 
\rangle\ll S$ and $\langle b^{\dagger}_{j}b_{j}\rangle\ll S$, followed by Fourier 
transformation
\begin{eqnarray}
a_{i} = \sqrt{\frac{2}{N}} \sum_{\mathbf{k}} e^{-i\mathbf{k} \cdot 
\mathbf{r}_{i}}a_{\mathbf{k}} \ \ \Leftrightarrow \ \ 
a_{\mathbf{k}}=\sqrt{\frac{2}{N}}\sum_{i}e^{i\mathbf{k}\cdot  
\mathbf{r}_{i}}a_{i} , 
\nonumber\\
b_{j}=\sqrt{\frac{2}{N}}\sum_{\mathbf{k}'}e^{-i\mathbf{k}' 
\cdot \mathbf{r}_{j}}b_{\mathbf{k'}} \ \ \Leftrightarrow \ \  
b_{\mathbf{k}'}=\sqrt{\frac{2}{N}}\sum_{j}e^{i\mathbf{k}' 
\cdot \mathbf{r}_{j}}b_{j} \nonumber\\
\end{eqnarray}
with orthogonality relations
\begin{eqnarray}
\sum_{i}e^{\pm i (\mathbf{k}-\mathbf{k}')\cdot 
\mathbf{r}_{i}}=\frac{N}{2}\delta_{\mathbf{k}\mathbf{k}'}, \ \ \   
\sum_{\mathbf{k}}e^{\pm i \mathbf{k}\cdot 
(\mathbf{r}_{j}-\mathbf{r}_{i})}=\frac{N}{2}\delta_{ij}.\nonumber\\ 
\end{eqnarray}
Here, $N$ is the number of sites and 
 \begin{eqnarray}
\gamma^{(1)}_{\mathbf{k}} = \frac{1}{z^{(1)}} \sum_{\boldsymbol{\delta}_{1}} 
e^{i\mathbf{k} \cdot \boldsymbol{\delta}_{1}},\ \ \ \gamma^{(2)}_{\mathbf{k}} = \frac{1}{z^{(2)}} \sum_{\boldsymbol{\delta}_{2}} 
e^{i\mathbf{k} \cdot \boldsymbol{\delta}_{2}} ,
\end{eqnarray}
where the sums are carried out over the vectors $\boldsymbol{\delta}_{1}$ connecting a magnetic site to its nearest neighbours and 
 the vectors $\boldsymbol{\delta}_{2}$ connecting a magnetic site to its next nearest neighbours. $z^{(1)}$ and $z^{(2)}$ are the numbers of nearest and next nearest neighbours of each site on the the lattice, respectively. The bosonized Hamiltonian in $\mathbf{k}$-space reads 
\begin{eqnarray}
H_{\mathbf{k}} & = & 
\omega_{a_{\mathbf{k}}}a_{\mathbf{k}}^{\dagger} a_{\mathbf{k}} + 
\omega_{b_{-\mathbf{k}}}b_{-\mathbf{k}}^{\dagger} b_{-\mathbf{k}}
\nonumber \\ 
 & & + g_{\mathbf{k}} a_{\mathbf{k}} b_{-\mathbf{k}} + g^{*}_{\mathbf{k}}a_{\mathbf{k}}^{\dagger}b_{-\mathbf{k}}^{\dagger}, 
\label{MHM-A}
\end{eqnarray}
where 
\begin{eqnarray}
\omega_{a_{\mathbf{k}}}&=&\epsilon_{\mathbf{k}}-B,\ \ \ \ \ \ 
\omega_{b_{-\mathbf{k}}}=\epsilon_{\mathbf{k}}+B,\nonumber\\
\epsilon_{\mathbf{k}}&=&S\left[z^{(1)}J^{(1)}+2\mathcal{K}\right.\nonumber\\
&&\left. +z^{(2)}\left(J^{(2)}-2\text{Re}[(J^{(2)}-iD^{(2)})\gamma^{(2)}_{\mathbf{k}}]\right)\right]\nonumber\\
g_{\mathbf{k}}&=&Sz^{(1)}\gamma^{(1)}_{\mathbf{k}}(J^{(1)}+iD^{(1)}).
\label{CCab}
\end{eqnarray}
Here, $a_{\mathbf{k}}^{\dagger}$ ($a_{\mathbf{k}}$) and $b^{\dagger}_{-\mathbf{k}}$ ($b_{-\mathbf{k}}$) are bosonic creation (annihilation) operators, which mutually commute and define independent families of bosonic operators on the opposite sublattices 
$A$ and $B$, respectively.

Under SU(1,1) Bogoliubov transformation
 \begin{eqnarray}
\left(
\begin{array}{cc}
  a_{\mathbf{k}}    \\
   b_{-\mathbf{k}}^{\dagger}       
\end{array}
\right)=\left(
\begin{array}{cc}
  u_{\mathbf{k}}& v_{\mathbf{k}}    \\
v_{\mathbf{k}}^{*}& u_{\mathbf{k}}^{*}       
\end{array}
\right)\left(
\begin{array}{cc}
  \alpha_{\mathbf{k}}    \\
   \beta_{-\mathbf{k}}^{\dagger}       
\end{array}
\right),
\label{eq:FBT-A}
\end{eqnarray}
where $u_{\mathbf{k}} =\cosh(r_{\mathbf{k}})$ and $v_{\mathbf{k}} = \sinh(r_{\mathbf{k}})e^{i\phi_{\mathbf{k}}}$ with
\begin{eqnarray}
r_{\mathbf{k}}&=&\tanh^{-1}\left[\frac{1-\sqrt{1-|\Gamma_{\mathbf{k}}|^{2}}}{|\Gamma_{\mathbf{k}}|}\right]\ge 0,\nonumber\\
\phi_{\mathbf{k}}&=&\pi-\arg[\Gamma_{\mathbf{k}}],\ \ \ \ 
\Gamma_{\mathbf{k}}=\frac{g_{\mathbf{k}}}{\epsilon_{\mathbf{k}}},
\end{eqnarray}
we obtain the following diagonal form of the Hamiltonian
 \begin{eqnarray}
H_{\mathbf{k}} = 
\omega_{\alpha_{\mathbf{k}}}\alpha_{\mathbf{k}}^{\dagger}\alpha_{\mathbf{k}} +
\omega_{\beta_{-\mathbf{k}}}\beta_{-\mathbf{k}}^{\dagger} \beta_{-\mathbf{k}}
\label{DHH-A}
\end{eqnarray}
with magnon dispersion relation 
\begin{eqnarray}
\omega_{\alpha_{\mathbf{k}}}&=&\tilde{\epsilon}_{\mathbf{k}}-B,\ \ \ \ \ \omega_{\beta_{-\mathbf{k}}}=\tilde{\epsilon}_{\mathbf{k}}+B\nonumber\\
\tilde{\epsilon}_{\mathbf{k}}&=&\cosh(2 r_{\mathbf{k}})\epsilon_{\mathbf{k}}+\sinh(2 r_{\mathbf{k}})\text{Re}(g_{\mathbf{k}}e^{i\phi_{\mathbf{k}}})
\end{eqnarray}
provided $|\Gamma_{\mathbf{k}}|<1$.
The SU(1,1) condition $|u_{\mathbf{k}}|^{2}-|v_{\mathbf{k}}|^{2}=1$
assures that $\alpha_{\mathbf{k}}$ and $\beta_{-\mathbf{k}}$ 
for all $\mathbf{k}$ also define independent families of bosonic operators.

\section{Description of magnon-magnon entanglement in terms of entanglement parameter}
\label{MME-long}
From the diagonal expression in Eq.~(\ref{DHH-A}), the ground state of 
the Hamiltonian $H_{\mathbf{k}}$ in the $(\alpha,\beta)$ mode reads
\begin{eqnarray}
\ket{\psi_{0}}= \ket{0; \alpha_{\mathbf{k}}}\ket{0; 
\beta_{-\mathbf{k}}},
\label{GS}
\end{eqnarray}
where $\ket{0; \alpha_{\mathbf{k}}}$ and $\ket{0; \beta_{-\mathbf{k}}}$ 
are vacuum states of $\alpha_{\mathbf{k}}$ and $\beta_{-\mathbf{k}}$, 
respectively, i.e.,  
\begin{eqnarray}
\alpha_{\mathbf{k}}\left(\ket{0; \alpha_{\mathbf{k}}}\ket{0; 
\beta_{-\mathbf{k}}}\right) & = &  \beta_{-\mathbf{k}}\left(\ket{0; 
\alpha_{\mathbf{k}}}\ket{0; \beta_{-\mathbf{k}}}\right) = 0. 
\nonumber \\ 
\label{eq:VEq}
\end{eqnarray}
By expressing the product vacuum state as a linear combination of $\ket{n; a_{\mathbf{k}}}$ and $\ket{n; b_{-\mathbf{k}}}$, which 
are the occupation number bases for the bosonic operators $a_{\mathbf{k}}$ and $b_{-\mathbf{k}}$, respectively, we find  
\begin{eqnarray}
\ket{0; \alpha_{\mathbf{k}}}\ket{0;\beta_{-\mathbf{k}}} = 
\sum_{n=0}^{\infty} p_{n; \mathbf{k}}\ket{n; a_{\mathbf{k}}} 
\ket{n; b_{-\mathbf{k}}},
\label{eq:SCS}
\end{eqnarray}
and by inserting it into Eq.~(\ref{eq:VEq}), we find 
\begin{eqnarray}
p_{n+1; \mathbf{k}}=\frac{v_{\mathbf{k}}}{u_{\mathbf{k}}^{*}} 
p_{n; \mathbf{k}},
\end{eqnarray}
where $v_{\mathbf{k}}$ and $u_{\mathbf{k}}$ are given by Bogoliubov transformation in Eq.~\eqref{eq:FBT-A}.
By solving this recursive equation with normalisation constraint 
$\sum_{n=0}^{\infty}\left| p_{n;{\bf k}} \right|^{2}=1$, the probability amplitudes in the superposed coherent state of Eq.~(\ref{eq:SCS}) 
become 
\begin{eqnarray}
p_{n; \mathbf{k}} = \frac{e^{in\phi_{\mathbf{k}}}}{\cosh 
r_{\mathbf{k}}}\tanh^{n}r_{\mathbf{k}}.
\label{EXCC}
\end{eqnarray}
Therefore, in the $(a,b)$ modes, the ground state magnetic modes can be written as a vector-valued function of $(r_{\mathbf{k}}, \phi_{\mathbf{k}})$:
\begin{eqnarray}
\ket{\psi_{0}(r_{\mathbf{k}}, \phi_{\mathbf{k}})}= \frac{1}{\cosh r_{\mathbf{k}}} 
\sum_{n=0}^{\infty} e^{in\phi_{\mathbf{k}}} \tanh^{n} r_{\mathbf{k}} 
\ket{n; a_{\mathbf{k}}}\ket{n; b_{-\mathbf{k}}}.
\nonumber\\
\label{eq:GSatab}
\end{eqnarray}
Note that we have in fact performed the inverse of the transformation in  
Eq.~(\ref{eq:FBT-A}) to derive Eq.~(\ref{eq:GSatab}) from Eq.~\eqref{GS}. 

Equations \eqref{GS} and \eqref{eq:GSatab} indicate that while 
the ground state is a product state in $(\alpha, \beta)$ modes, it is an entangled state when expressed in $(a, b)$ modes with the entanglement entropy given by
\begin{eqnarray}
E\left[\ket{\psi_{0}(r_{\mathbf{k}}, \phi_{\mathbf{k}})}\right] & = & \left[ \cosh^{2} (r_{\mathbf{k}}) \log \cosh^{2} (r_{\mathbf{k}}) \right. 
\nonumber\\
& & \left. -\sinh^{2}(r_{\mathbf{k}})\log\sinh^{2}(r_{\mathbf{k}})\right].
\end{eqnarray}
A hierarchy of magnon mode entanglement in antiferromagnets has been studied in Ref.~\cite{azimi-mousolou2020}.
Note that the entanglement is quantified by the  parameter $r_{\mathbf{k}}$, or equivalently $|\Gamma_{\mathbf{k}}|$. This is indeed true for the complete energy eigenbasis states, which take the form $(\alpha^{\dagger}_{\mathbf{k}})^{x}(\beta^{\dagger}_{-\mathbf{k}})^{y}\ket{\psi_{0}}=\sqrt{x!y!}\ket{x; \alpha_{\mathbf{k}}}\ket{y; \beta_{-\mathbf{k}}}$ for any positive integer powers of $x$ and $y$ that, although each of them is disentangled in the $(\alpha, \beta)$ modes, are entangled states in the $(a, b)$ magnon modes with an entanglement entropy that only depends on the parameter $r_{\mathbf{k}}$, or equivalently $|\Gamma_{\mathbf{k}}|$. To see this, we note that  
\begin{eqnarray}
(\alpha^{\dagger}_{\mathbf{k}})^{x}(\beta^{\dagger}_{-\mathbf{k}})^{y}\ket{\psi_{0}}&\equiv&
\sum_{n=0}^{\infty} p^{(x, y)}_{n; \mathbf{k}}\ket{n+\delta m; a_{\mathbf{k}}} 
\ket{n; b_{-\mathbf{k}}},\ x\ge y,
\nonumber\\
(\alpha^{\dagger}_{\mathbf{k}})^{x}(\beta^{\dagger}_{-\mathbf{k}})^{y}\ket{\psi_{0}}&\equiv&
\sum_{n=0}^{\infty} p^{(x, y)}_{n; \mathbf{k}}\ket{n; a_{\mathbf{k}}} 
\ket{n+\delta m; b_{-\mathbf{k}}},\ x\le y,
\nonumber\\
\label{EES}
\end{eqnarray}
by induction and $\delta m=|x-y|$. Here, the probability amplitudes are given by
\begin{eqnarray}
p^{(x, y)}_{n; \mathbf{k}}=\frac{1}{\sqrt{x!y!}}
\left(\frac{1}{u_{\mathbf{k}}^{*}}\right)^{\delta m}\left(\frac{1}{u_{\mathbf{k}}^{*}v_{\mathbf{k}}}\right)^{m}q^{(m, \delta m)}_{n; \mathbf{k}}p_{n; \mathbf{k}},\ \ \ 
\end{eqnarray}
where $m=\min\{x, y\}$ and $p_{n; \mathbf{k}}$ is the expansion coefficient given in Eq.~\eqref{EXCC}. The $q^{(m, \delta m)}_{n; \mathbf{k}}$ satisfies the following recursive relations 
\begin{eqnarray}
q^{(m, \delta m>0)}_{n; \mathbf{k}}&=&|u_{\mathbf{k}}|^{2}\sqrt{n+\delta m}q^{(m, \delta m-1)}_{n; \mathbf{k}}
\nonumber \\  & & -|v_{\mathbf{k}}|^{2}\sqrt{n+1}q^{(m, \delta m-1)}_{n+1; \mathbf{k}} , \nonumber\\
q^{(m>0, 0)}_{n; \mathbf{k}}&=&n|u_{\mathbf{k}}|^{4}q^{(m-1, 0)}_{n-1; \mathbf{k}}-(2n+1)|u_{\mathbf{k}}v_{\mathbf{k}}|^{2}q^{(m-1,0)}_{n; \mathbf{k}}\nonumber\\
&&+(n+1)|v_{\mathbf{k}}|^{4}q^{(m-1,0)}_{n+1; \mathbf{k}},
\end{eqnarray}
with initial value condition $q^{(0, 0)}_{n; \mathbf{k}}=1$ for each $n$.

Since $|u_{\mathbf{k}}|$, $|v_{\mathbf{k}}|$, and the probabilities $|p_{n; \mathbf{k}}|^{2}$ are only functions of $r_{\mathbf{k}}$, the probabilities $|p^{(x, y)}_{n; \mathbf{k}}|^{2}$ are only functions of $r_{\mathbf{k}}$. This demonstrates that the entanglement entropies of all energy eigenbasis states in the $(a, b)$ modes, i.e.,
\begin{eqnarray}
E\left[(\alpha^{\dagger}_{\mathbf{k}})^{x}(\beta^{\dagger}_{-\mathbf{k}})^{y}\ket{\psi_{0}}\right]&=&-\sum_{n=0}^{\infty}|p^{(x, y)}_{n; \mathbf{k}}|^{2}\log|p^{(x, y)}_{n; \mathbf{k}}|^{2}\nonumber\\
\label{EEE-A}
\end{eqnarray}
are only functions of the single squeezing parameter $r_{\mathbf{k}}$.
For this reason we identify $r_{\mathbf{k}}$, or equivalently $|\Gamma_{\mathbf{k}}|$, as the magnon-magnon entanglement parameter of the system.

Note that $H_{1}=0$ corresponds to  $r_{\mathbf{k}}=0$ and trivial Bogoliubov transformation in Eq.~\eqref{eq:FBT-A}. In this case the $(a, b)$ and $(\alpha, \beta)$ modes coincide and therefore both Hamiltonians in Eqs.~\eqref{MHM} and \eqref{DHH} describe two separate non-interacting ferromagnets with opposite magnetizations. Moreover, the full energy spectra are given by product states  $\ket{n; a_{\mathbf{k}}}\ket{m; b_{-\mathbf{k}}}$, which indicate $n$ and $m$ ferromagnetic magnons in the bosonic modes of $a\equiv\alpha$ and $b\equiv\beta$ respectively. However, for $H_{1}\ne 0$,  the nontrivial transformation in Eq.~\eqref{eq:FBT-A} combines the two oppositely magnetized ferromagnetic magnon modes of  $a$ and  $b$, which represent two ferromagnetic sublattices, into antiferromagnetic magnon modes $(\alpha, \beta)$. Thus, in the presence of antiferromagnetic interaction $H_{1}$,  the trivial entanglement entropy between $\alpha$ and $\beta$ modes means no entanglement between antiferromagnetic magnons in the system while non-zero entanglement entropy in $(a, b)$ modes indicate entanglement between the two ferromagnetic magnons or the two ferromagnetic sublattices.

\end{document}